# Dynamical Anisotropy of a Colloidal Glass Under Pressure

Shengyun Shi[1], Li Tian[2, †] and Bo Li[1, *]

1 The Institute for Advanced Studies, Wuhan University, Wuhan 430072, P. R. China.

2 Department of Physics, Wenzhou University, Wenzhou, 325000, P.R. China

Corresponding author names: Bo Li, Li Tian

**Email:** * libo@whu.edu.cn, † li.tian@wzu.edu.cn

**Author Contributions:** B.L. and L.T. designed the research. S.Y.S. performed the experiment and analyzed the data. All authors discussed the results and wrote the paper.

**Competing Interest Statement:** The authors declare no competing interests.

**Classification:** Physical Sciences; Materials Physics

**This PDF file includes:**

Main Text
Figures 1 to 4

## Abstract

Pressure is a critical thermodynamic parameter that profoundly influences the physical properties of glasses. Pressure-induced densification and structural transformation endow glasses manufactured under such conditions with exceptional mechanical and optical properties. Although pressure-treated glasses have been characterized by ensemble-averaged methods such as X-ray diffraction and Raman spectroscopy, their microscopic dynamics have rarely been addressed, limiting our understanding of the coupling among structure, dynamics, and mechanics. Here, using a binary hard-sphere colloidal glass confined in cylindrical capillary tubes, we impose a constant pressure on the particles through the tangential component of gravity. This approach enables the first investigation of pressure effects on colloidal glasses. We find that the dynamics are substantially frozen, whereas the structural change remains comparatively mild. At low pressure, spatial correlations among structural, dynamical, and local elastic heterogeneities are observed. Remarkably, dynamical anisotropy emerges in response to pressure, characterized by faster motion parallel to the pressure direction than perpendicular to it. This anisotropy is attributed to an instability induced by the strong pressure force. Concurrently, structural and dynamical heterogeneities are strongly suppressed under pressure. Our experiments characterize the microscopic dynamics of colloidal glasses under pressure and offer design principles for the manufacturing protocol of glass materials.

## Significance Statement

Pressure is widely used to fabricate glasses with exceptional properties, yet its influence on microscopic dynamics remains largely unexplored. Using a colloidal glass subjected to constant pressure, we directly visualize particle motion and uncover a striking dynamical anisotropy: particles move faster parallel to the pressure direction than perpendicular to it, an effect linked to pressure-induced instability. Pressure also freezes dynamics, homogenizes local structure, and suppresses temporal variation. These findings provide a microscopic picture of glasses under pressure and offer design principles for advanced glass materials.

## Introduction

Glass materials are widely used in modern technology for their optical and mechanical properties (1). Pressure not only affects the stable-state properties of glasses but also alters the kinetic pathways (2–4). Structural analyses have revealed pressure-induced densification (5–7), amorphous–amorphous transformations (8), crystallization (9), and vitrification (10, 11) in glass materials, typically accompanied by local structural rearrangements (12-18). The pressure–temperature ($P$–$T$) phase diagram (19) encompasses a broader glass region than the temperature–density ($T$–$\rho$) diagram. Consequently, imposing high pressure during classical thermal cooling has yielded materials with substantially altered physical properties (20-24). However, conventional techniques such as X-ray diffraction, Raman spectroscopy, and neutron scattering lack single-particle resolution and discrete time windows, providing only ensemble-averaged structural information and leaving microscopic dynamics under pressure largely unexplored.

Colloids are outstanding model systems for studying phase behavior in atomic or molecular materials, offering access to fast kinetics and microscopic information (25–27). Combined with video microscopy (28), colloidal systems have been widely used to reveal evolving length scales (29), mechanical strength origins (30), and dynamical heterogeneity (31) in glasses. Beyond natural packing, confinement and curvature effects have been investigated in colloidal glasses (32–35) and crystals (36–41), showing that geometry influences both structure and dynamics. However, pressure remains much less explored in colloidal systems because it is challenging to impose and vary hydrostatic pressure quantitatively. Monte Carlo simulations have examined dynamical heterogeneity of supercooled liquids under sudden compression (42), resembling indentation or impact experiments on real glasses, but have not addressed the deep glass region under sustained pressure. Theoretical studies have proposed *P–T* phase diagrams within mode-coupling theory (43), and molecular dynamics simulations have analyzed local structure under pressure (44, 45). Nevertheless, none has investigated dynamics with single-particle resolution. Thus, experimental realization of colloidal glasses under pressure and the underlying correlation among microscopic dynamics, structure, and elasticity remain open questions.

To address these questions, we developed a method to impose lateral pressure (*P*) on quasi-two-dimensional colloidal glasses. As illustrated in **Fig. 1A**, binary silica particles (2 μm and 3 μm, number ratio 1:1, following the Kob–Andersen model (46) to suppress crystallization) are injected into cylindrical capillary tubes. Owing to the significant density difference with water, particles sediment to the tube bottom, forming a quasi-two-dimensional observation region ("Observational Zone in **Fig. 1A, B**"). The two flanking regions ("Pressure Zone in **Fig. 1A, B**") sandwiching the observation region compress it into a deep glassy state, serving as the pressure source (**Fig. 1B**), analogous to the classical diamond anvil cell experiments in atomic glasses (47). By tuning particle concentration, the flanking region width is quantitatively controlled, yielding different pressure strengths (**Table S1**). Specifically, higher particle concentration leads to a wider flanking region, the tangential component of which eventually provides larger pressure to the observational zone sedimented at the bottom of the capillary tube. The sample was mounted on an epi-illumination microscope, and video microscopy was performed at 28.85, 250, and 2000 frames per second (fps). Details on sample preparation, pressure quantification (**Eqs. 1–6**), and data analysis (**Eqs. 7–17**) are provided in **Methods**.

Using this platform, we systematically investigated five pressures (**Fig. 1C**) and compared results with a classical two-dimensional model varying area fraction ($\varphi$) without applied pressure. To our knowledge, we provide the first revelation of how pressure influences colloidal glass properties by resolving microscopic dynamics at the single-particle level. Under pressure, dynamics and relaxation slow dramatically (**Fig. 1D–F**), accompanied by structural densification and enhanced mechanical strength (**Fig. 2A–E**). Coupling among structure, dynamics, and elasticity is observed at low pressure but disappears at high pressure (**Fig. 2F, G**). Surprisingly, significant dynamical anisotropy emerges on ultrafast timescales, with faster motion parallel to pressure (**Fig. 3**). In contrast, pressure homogenizes structure and suppresses variation with time (**Fig. 4**). The observations and trend are consistent for all five curvatures ($\kappa$), proving our findings are caused solely by pressure.

## Results

### Pressure Dramatically Slows Glass Dynamics.

To characterize the dynamics, we calculated the mean square displacement (*MSD*, **Eq. 7**) (**Fig. 1D–F**) and the intermediate scattering function ($F_s(q, t)$**, Eq. 8**) (**Fig. 1G–I**). Both quantities exhibit behavior characteristic of deep glasses, characterized by the plateaus in *MSD* (**Fig. 1D, E**) and two-step relaxation in $F_s(q, t)$ (**Fig. 1G, H**). Even at mild pressure, plateaus appear in the *MSD* curves (**Fig. 1D, E**), indicating that particle motion is caged by neighboring particles. When *P* reaches $4.56 \times 10^{-6}$ Pa, the *MSD* remains constant over the largest accessible timescales. The degree of dynamical suppression induced by pressure is remarkable because, in a two-dimensional hard-sphere system, the *MSD* typically increases again after the plateau owing to long-wavelength fluctuations (48). The $F_s(q, t)$ curves display clear plateaus and two-step relaxation (**Fig. 1G, H**), hallmarks of supercooled liquids and glasses (1). Consistent with the *MSD* data, $F_s(q, t)$ does not decay when the pressure is sufficiently high. Larger curvature leads to stronger dynamical suppression (**Fig. 1D, E, G, H**) because the effective pressure component of gravity increases. To better illustrate this dynamical suppression, we performed classical two-dimensional binary glass experiments in which the area fraction was varied using the same silica particles. Overall, the dynamics and relaxation of glasses under pressure (**Fig. 1D, E, G, H**) are comparable to those of a system with $\varphi = 0.83$ on a flat surface without pressure (**Fig. 1F, I**).

To quantify the degree of dynamical slowing as a function of pressure, we averaged the particle displacement within $\Delta t = 100$ s (**Fig. 2A**), the time at which the *MSD* reaches a plateau (**Fig. 1D, E**). For all five curvatures $\kappa$, the displacement decays faster than linearly with $P$, yielding an order-of-magnitude decrease. In contrast, the accompanying structural densification and mechanical enhancement are much milder. The differences in coordination number (characterizing local neighborhood configuration) (**Fig. 2B**) and Voronoi cell area ($A_{vor}$, characterizing cage size, **Eq. 9**) (**Fig. 2C**) between the glasses under the smallest and largest pressures are less than 10%, without any observable structural phase transformation. Notably, in our flat samples where only $\varphi$ is varied, an order-of-magnitude decrease in dynamics requires a reduction in Voronoi cell area of at least 35%. Therefore, the dramatic dynamical slowdown induced by pressure cannot be explained solely by structural caging from neighboring particles. The estimated elastic modulus ($G$, **Eq. 10**) increases linearly by two- to four-fold with $P$, depending on $\kappa$ (**Fig. 2D**), a trend also far milder than that of the dynamics. This is reasonable because, in essence, the mechanical strength of a material is determined by its microscopic structure.

**Spatial Correlation Between Dynamics, Structure, and Mechanics.**

Since dynamical slowing (**Fig. 2A**), structural densification (**Fig. 2B, C**), and mechanical hardening (**Fig. 2D**) occur simultaneously in response to pressure, we ask whether these phenomena are spatially correlated. Plotting G as a function of Avor, we find that all data—regardless of $P$, $\kappa$, and $\varphi$—collapse onto a universal master curve without any renormalization (**Fig. 2E**). This provides strong evidence that the mechanical strength of the glass is governed by its structure, consistent with the basic scenario of elasticity theory in solid materials (49, 50).

Fitting yields an exponential relationship between $G$ and $A_{vor}$, with a decay exponent of −1.08. The data can also be described by a power-law decay, as they likewise collapse onto a straight line on a log-log plot. The fitted decay power is one order of magnitude larger than the −1.5 predicted by theory. This trend departs from the theoretical prediction of power-law decay in bulk glasses (51), which spectroscopy experiments on silica glass have qualitatively validated (52, 53). In fact, those experiments also report a larger decay power than the predicted value. In short, $G$ decays with $A_{vor}$ considerably faster than the power law predicted by theory or measured in three-dimensional bulk glasses. Therefore, in this two-dimensional system, mechanical strength is far more sensitive to structure than the theoretical prediction: a very slight decrease in local free volume leads to dramatic hardening, presumably because the reduced degrees of freedom in two dimensions leave particles with less space to relax. Although a power-law dependence of $G$ on $A_{vor}$ cannot be excluded, the fact that the trend is even closer to exponential decay, together with the large fitted decay power, challenges one or more basic theoretical assumptions of affine deformation, pairwise interactions, or free-volume scaling.

The effect of pressure on spatial coupling is intriguing. At relatively mild pressure, spatial mapping among displacement, coordination number, $A_{vor}$, local elasticity components ($\sigma_{xx}$, $\sigma_{xy}$, $\sigma_{yy}$) (54), and the six-fold orientational order parameter ($\psi_6$, **Eq. 11**) can be observed (**Fig. 2F**). The local regions that map the best - for example, the region in the circle of **Fig. 2F** - usually exhibit signs of pressure-induced ordering at some local spots. This rationalizes the observed spatial mapping at small $P$ because, as the global energy minimum, crystalline order naturally corresponds to slower dynamics and less stress. At high pressure, however, such spatial mapping does not exist (**Fig. 2G**). Meanwhile, local crystalline order also disappears, indicating that the system is completely amorphized again. The spatial mapping and its intriguing trend distinguish pressure from other thermodynamic variables such as temperature and density. In thermally cooled glasses, such clear spatial mapping among structure, elasticity, and dynamics has not been observed. In the classical scenario where viscosity increases upon cooling, dynamical heterogeneity emerges near the mode-coupling temperature $T_c$, and its associated timescales and length scales increase monotonically as temperature decreases toward the ideal glass transition temperature $T_g$ (55). In contrast, the complex evolution of spatial coupling observed here, together with the results presented in the following sections, indicates that pressure plays a more intricate role in the relaxation modes of glass materials than temperature in thermal processes.

**Pressure-Induced Ultrafast Dynamical Anisotropy.**

Surprisingly, dynamical anisotropy emerges on an ultrafast timescale. Using a high-speed camera, we increased the image capture rate to 2000 fps, corresponding to a time resolution of $5 \times 10^{-4}$ s. In the presence of pressure, the displacement parallel to the pressure direction (*x*-direction) becomes significantly larger than that perpendicular to it (*y*-direction) when the time interval $\Delta t$ is on the order of one second or less (**Fig. 3A–D**). This is an unambiguous signature of dynamical anisotropy, as the *x*- and *y*-directions are no longer equivalent. The difference increases as $\Delta t$ decreases to faster timescales but converges to zero at large timescales (**Fig. 3A–D**). The displacement difference exists for all $\kappa$ and has not been reported in any glass system controlled by temperature, indicating that this dynamical anisotropy is a direct consequence of pressure. The probability distributions of displacement in the *x*- and *y*-directions confirm the anisotropy and, in addition, rule out non-Gaussian behavior (56) at such short timescales (**Fig. 3E–H**).

To characterize the microscopic features of the dynamical anisotropy, we plotted displacement vectors at selected $\Delta t$, colored according to direction (**Fig. 3I–L**). For all $P$ and $\kappa$ data at different $\Delta t$, we did not observe apparent cooperative rearrangement regions (57, 58); the spatial distribution of vector magnitudes is rather sporadic. The vectors tend to align along the *x*-direction, but they are neither systematically larger in magnitude nor spatially clustered. Therefore, the dynamical anisotropy is unlikely to arise from the dynamical heterogeneity that typically accompanies thermally cooled glasses. At small $\Delta t$, however, more vectors align along the *x*-direction, indicating the particle movement is biased along this direction. This real-space observation is confirmed by the probability distribution of displacement orientation (**Fig. 3M–P**). At large $\Delta t$, however, the distribution becomes isotropic. Hence, it is the direction rather than the magnitude of particle motion that underlies the dynamical anisotropy.

Notably, the probabilities of displacement in the $+x$ and $-x$ directions show no discernible difference, ruling out sudden collective motion in a single direction as the origin. The area fraction remains constant throughout the experiments, confirming that the dynamical anisotropy is a stable-state phenomenon rather than an emergent collective dynamics driven by pressure-induced aging (59). There exists no gradient in the spatial distribution of any structural, elastic, or dynamical quantity along the pressure direction (**Fig. 2F, G**), proving that the pressure forces from both flanking sides are well balanced, and thus the stress is uniformly distributed across the observational zone. This excludes another driven collective motion caused by a transient unbalanced force that could possibly have caused the dynamical anisotropy.

It is worth emphasizing that the dynamical anisotropy occurs at very short timescales. Significant differences between displacements along the *x*- and *y*-directions appear in the time range from $10^{-3}$ s to 1 s **(Fig. 3A–D**). Based on the $F_s(q, t)$ data (**Fig. 1G, H**), the $\beta$-relaxation time of this system lies in the range of 10–40 s, while the $\alpha$-relaxation time exceeds 1000 s. Thus, when the dynamical anisotropy occurs, the particles are still well inside their cages, undergoing rattling motion around their equilibrium positions. As noted in the $G$–$A_{vor}$ relation reported in **Fig. 2E**, the mechanical strength is extremely sensitive to free volume. This sensitivity could reasonably also induce instability in the dynamics, because tiny inhomogeneities in local stress arising from structural change may bias the dynamics at short times or small length scales, thereby offering a plausible explanation for the observed dynamical anisotropy.

**Suppressed Heterogeneity and Temporal Variation Under Pressure.**

In sharp contrast to the dynamical anisotropy reported above, the system under pressure is in fact highly homogeneous and quiescent when assessed from the perspectives of structure and variation in time. To quantify the structure, we calculated the radial distribution function ($g(r)$, **Eq. 12**), the geodesic distribution function ($g(s)$, **Eq. 13**), and the angle distribution function ($g(\Delta\theta)$, **Eq. 14**) (**Fig. 4A–D**), and plotted their peak heights (**Fig. 4E–H**) and positions (**Fig. 4I–L**) as functions of $P$. Compared with the situation on a flat surface with varying $\varphi$, all three quantities are insensitive to $P$, indicating little structural change. In real space, the distributions of $A_{vor}$ and coordination number are highly homogeneous, with no evidence of crystallization (**Fig. 2F, G**). The shape of the Voronoi cells is overall isotropic, with no sign of compression along the pressure direction. As shown in **Fig. 1D–I**, even slight pressure is sufficient to push the system into the deep glass region, where particles are closely packed. The reduced space available for particle relaxation or cooperative motion traps the system in a deep local minimum of the free-energy landscape, corresponding to a homogeneous structure (60). Unlike temperature, the effect of $P$ appears to act in an on–off manner: once $P$ exceeds a threshold value, the resulting structure remains essentially unchanged.

Time variations are likewise suppressed by pressure. The susceptibility of $A_{vor}$, $\chi_{area}$ (**Eq. 15**), decreases with $P$ (**Fig. 4M**), indicating that even cage rattling is reduced, freezing the system into a highly quiescent state. To quantify dynamical heterogeneity, we further calculated the four-point correlation function ($\chi_4$, **Eqs. 16 and 17**) (**Fig. 4N**). The peak height of $\chi_4$ unexpectedly decreases with $P$ (**Fig. 4O**), demonstrating that the system becomes less dynamically heterogeneous. The smaller correlation length suggests less cooperative rearrangement. This is consistent with real-space observation (**Figs. 2F, G, 3I-L**)—particle displacements are sporadic with little evidence of collectiveness. The strong pressure confines all particles inside the cages formed by their nearest neighbors, leaves no space for cooperative relaxation, and therefore eliminates any long-range correlation. In systems controlled by temperature, dynamical heterogeneity emerges when the system is cooled below $T_c$, with cooperative length scales increasing monotonically (55). A strict decrease in dynamical heterogeneity and the associated length scales with a thermodynamic variable—here $P$—has not been observed. Therefore, although both cooling (lower $T$) and pressurization (higher $P$) freeze the system into less dynamical states, they likely follow fundamentally different kinetic pathways that lead to distinct local minima in the free-energy landscape. Under high $P$, the system attains a unique state characterized by uniform and insensitive structure, suppressed temporal variation and dynamical heterogeneity, and slow but anisotropic dynamics. The peak position of $\chi_4$ increases with $P$, indicating an increasing timescale that also occurs in cooling glasses (55).

## Discussion

In this work, we developed a flanking gravity method to impose constant pressure on quasi-two-dimensional colloidal glasses and investigated their physical properties—dynamics, structure, and mechanics—under such conditions that have not, to our knowledge, been accessed experimentally. Pressure dramatically slows the system, homogenizes the structure, and suppresses variation in time. However, several more intriguing findings emerge from our experiments: the spatial correlation among dynamics, structure, and mechanics evolves intriguingly with $P$; dynamical anisotropy appears at short timescales; and dynamical heterogeneity decreases with pressure. These observations fundamentally distinguish pressurized glasses from temperature-controlled glasses obtained through thermal cooling.

One open question raised by our experiments, relevant to both theorists and experimentalists, is the origin of the short-time dynamical anisotropy—what causes the asymmetry between the directions parallel and perpendicular to the external pressure? More generally, why do anisotropy and instability emerge in a system where heterogeneity and variation are largely suppressed by pressure? Although we have shown that the structure is highly uniform under pressure, there may be subtle structural features that could be captured by more sophisticated order parameters (45, 61, 62). This assumption is rationalized by the extreme sensitivity of mechanical strength to free-volume change in this system (**Fig. 2E**), but it nonetheless requires more detailed validation. Another possibility is that pressure influences high-frequency vibrational modes, whereas most current studies focus on the low-frequency region (63). Regarding the correlation among dynamics, structure, and mechanics, more detailed spatial analysis is required to further confirm its connection with the local disorder-order-disorder transformation with increasing $P$. Although we have focused on pressure, our experiments also systematically probe colloidal glasses on curved surfaces with different curvatures, an effect highly relevant to the surface and mechanical properties of glasses that deserves further exploration. Looking ahead, our capillary tube method enables the assembly of other types of materials, such as crystals (64) and liquid crystals (65), under pressure or on curved surfaces, facilitating not only mechanistic studies but also the fabrication of functional materials.

## Materials and Methods

**Sample preparation.** Carboxyl-functionalized silica microspheres (2 μm, 2.5% w/v, loose density 0.47 g/cm$^3$, Macklin, product No. M875489) and 3 μm particles of the same series were mixed at a 1:1 number ratio and dispersed in 0.3 wt% SDS aqueous solution (deionized water). Two types of samples were prepared. For confined samples (capillary group), 10 μL each of the 2 μm and 3 μm stock suspensions were mixed and diluted with 0.3 wt% SDS aqueous solution to a particle number concentration of approximately $10^7$–$10^8$ particles/mL. One end of the capillary was dipped into the suspension, and the liquid was drawn into the tube by capillary action. Glass capillaries with inner diameters of 0.3, 0.4, 0.5, 0.6, and 0.9 mm were used. The capillary was secured in the slide holder, with its ends fixed by Blu-Tack and the holder screws tightened to prevent any movement during the experiment. The assembly was placed on the microscope stage and equilibrated for at least 1 h before data acquisition. In this group, the particle layer within the ROI reached a locally saturated density; therefore, the state of the system is characterized by the lateral pressure $P$, rather than by volume fraction. For control samples (flat group), a 1 μL droplet of the suspension was deposited onto a glass coverslip and covered with a coverslip. The edges were sealed with AB glue. After the glue solidified, the sample was equilibrated on the microscope stage for at least 1 h before measurement. Six volume fractions were prepared for this group ($\varphi$ = 0.53, 0.59, 0.65, 0.71, 0.77, 0.83).

**Video microscopy and data correction.** Imaging was performed on an inverted microscope (Nexcope NIB610) equipped with PlanF S-Apo objectives: 20×/0.45 (air), 40×/0.75 (air), and 100×/1.45 (oil immersion). Bright-field illumination with white light was used throughout. The pixel calibrations for the three objectives are: 20×: 10 μm = 41.40 pixel, 40×: 10 μm = 92.47 pixel, and 100×: 10 μm = 204.80 pixel. Two cameras were employed. A high-speed camera (SH6-504-M-80, SSZN) was used at 2000 fps and 250 fps. A CMOS camera (UI-3180CP-C-HQ Rev.2.1, IDS Imaging Development Systems GmbH, resolution 2592 × 2048) was used at 28.85 fps for 2 h per experiment. Data from the three frame rates were concatenated to construct the full *MSD* curves (**Fig. 3A–D**). Image processing was performed using custom MATLAB routines with the Crocker–Grier tracking algorithm(28).

For capillary samples, the Z-coordinate was calculated as:

$$Z = -\sqrt{R^2 - (x - x_c)^2} \quad [1]$$

where $R$ (in pixel) is the capillary inner radius, $x_c$ is the pixel coordinate of the tube center, and $x$ is the horizontal pixel coordinate of the particle. The x-coordinate was then unfolded to arc length

$$s = R \arcsin(x/R) \quad [2]$$

to correct for projection distortion.

**Pressure quantification.** To quantify the lateral confinement pressure on particles adsorbed on the capillary inner wall, we developed a model based on geometric relations and force balance. For a capillary of radius $R$, the chord width $w$ in top-view projection relates to the central angle $\alpha$ as $w = 2R\cdot\sin\alpha$, giving $\alpha = \arcsin(w/2R)$. Let $w_t$ and $w_v$ be the chord widths of the total particle layer and the ROI, with corresponding half-angles $\alpha_t$ and $\alpha_v$.

Along the arc direction, the compressive force per unit axial length $T(\theta)$ satisfies the force balance. The position of a particle on the tube wall is described by the central angle $\theta$ (with $\theta = 0$ at the bottom center), and the tangential component of gravity is $mg\cdot\sin\theta$. The force balance equation is:

$$\frac{dT}{d\theta} = -n_s m_{\text{avg}} gRL \sin\theta \quad [3]$$

where $n_s$ is the areal number density, $m_{avg}$ is the average mass per particle, $g$ is the gravitational acceleration, and $L$ is the axial length. With the boundary condition $T(\alpha_t) = 0$ (no support at the outermost edge), integration gives:

$$T(\theta) = n_s m_{\text{avg}} gRL(\cos\theta - \cos\alpha_t) \quad [4]$$

The measured particle number in the ROI is $N_v$, giving $n_s = N_v / (2\alpha_v\cdot R\cdot L)$. Substituting yields the compressive force per unit axial length at the ROI boundary:

$$T(\alpha_v) = \frac{N_v m_{\text{avg}} g}{2\alpha_v}(\cos\alpha_v - \cos\alpha_t) \quad [5]$$

Multiplying $T(\alpha_v)$ by the axial length $L$ gives the total force, which is then divided by the projected area $w_v\cdot L$ . Using $\cos\alpha = \sqrt{1-\left(\frac{w}{2R}\right)^2}$ , the average lateral pressure is:

$$p = \frac{N_v m_{\text{avg}} g}{2\alpha_v w_v}\left(\sqrt{1-\left(\frac{w_v}{2R}\right)^2} - \sqrt{1-\left(\frac{w_t}{2R}\right)^2}\right) \quad [6]$$

where $w_v$ is the ROI width, and $w_t$ is the total particle layer width. Five pressure levels were defined: 35.6, 17.8, 11.2, 4.56, and 3.37 μPa. The corresponding $w_t$ values for each tube diameter are listed in **Table S1**.

**Table S1.** Particle layer total width $w_t$ (μm) for each tube diameter and pressure level.

| p (μPa) | 0.3 mm | 0.4 mm | 0.5 mm | 0.6 mm | 0.9 mm |
|---|---|---|---|---|---|
| 35.6 | 211.1 | 244.0 | 269.34 | 302.52 | 351.58 |
| 17.8 | 159.0 | 180.8 | 197.47 | 220.49 | 253.46 |
| 11.2 | 131.9 | 148.4 | 161.28 | 179.23 | 204.78 |
| 4.56 | 94.1 | 103.4 | 111.12 | 121.88 | 137.14 |
| 3.37 | 85.2 | 92.9 | 99.10 | 108.04 | 120.74 |

Note: $w_t$ is the total chord width of the particle layer measured in top-view projection. $w_v$ (ROI width) remains constant at 58.59 µm for all capillary samples. All values are in micrometers (µm).

**Physical properties characterization.** The following quantities were computed in the order they appear in the main figures. The mean square displacement was computed as:

$$\langle \Delta r^2(\tau) \rangle = \left\langle \left| \mathbf{r}(t+\tau) - \mathbf{r}(t) \right|^2 \right\rangle \quad [7]$$

where $\tau$ is the lag time and $\langle \cdot \rangle$ denotes averaging over all particles and starting times.

The self-intermediate scattering function was computed at the wave vector $q = 2\pi / r_1$, where $r_1$ is the first peak position of the radial distribution function $g(r)$, as:

$$F_s(q,t) = \left\langle \frac{1}{N} \sum_{j=1}^{N} e^{i\mathbf{q}\cdot[\mathbf{r}_j(t) - \mathbf{r}_j(0)]} \right\rangle \quad [8]$$

Voronoi tessellation was performed on the particle coordinates. For each particle $i$, the Voronoi cell $V_i$ is defined as:

$$V_i = \{ \mathbf{x} \in \mathbb{R}^2 \mid \| \mathbf{x} - \mathbf{r}_i \| \le \| \mathbf{x} - \mathbf{r}_j \|, \forall j \ne i \} \quad [9]$$

where $\boldsymbol{r}_i$ is the position vector of particle $i$. The cell area $A_i^{\text{proj}}$ was computed via the shoelace formula and corrected for projection distortion on the cylindrical wall as $A_i^{\text{real}} = \frac{A_i^{\text{proj}}}{\cos\theta_i}$, where $\theta_i = \arcsin(x_i / R)$.

The shear modulus was estimated as:

$$G = \frac{k_B T}{\langle \Delta r^2 \rangle} \cdot R \quad [10]$$

where $R$ is the average particle radius and $\langle \Delta r^2 \rangle$ is the *MSD* plateau height.

The coordination number for each particle was computed as $Z_i = \sum_{j \ne i} \Theta(r_{\text{cut}} - r_{ij})$, where $r_{\text{cut}} = 1.4 \cdot d_{\text{avg}}$ is the cutoff distance and $d_{\text{avg}}$ is the average particle diameter.

The sixfold orientational order parameter was computed as:

$$\psi_6(i) = \left| \frac{1}{N_i} \sum_{j=1}^{N_i} e^{i6\theta_{ij}} \right| \quad [11]$$

where $N_i$ is the number of nearest neighbors of particle $i$ and $\theta_{ij}$ is the angle between the vector connecting particle $i$ and its neighbor $j$ and a fixed reference axis. $\psi_6$ ranges from 0 to 1, with 1 indicating perfect sixfold order and 0 indicating complete disorder.

The radial distribution function was computed as:

$$g(r) = \frac{1}{\rho} \left\langle \sum_{j \ne i} \delta(r - r_{ij}) \right\rangle \quad [12]$$

The geodesic counterpart $g(s)$ was obtained from $s = R \arcsin(x/\mathrm{R})$, with pair distances defined as:

$$s_{ij} = \sqrt{(\Delta s)^2 + (\Delta y)^2} \quad [13]$$

The angular pair distribution was computed as:

$$g(\Delta\theta) = \left\langle \sum_{j \ne i} \delta(\Delta\theta - |\theta_i - \theta_j|) \right\rangle \quad [14]$$

where $\theta_i = \arcsin(x_i/\mathrm{R})$.

The area compressibility was defined as:

$$\chi_{\text{area}} = A \cdot \frac{\mathrm{Var}(a)}{\langle a \rangle^2} \quad [15]$$

where $a$ is the Voronoi area of a single particle, $A$ is the total ROI area, and $\mathrm{Var}(a) = \langle a^2 \rangle - \langle a \rangle^2$ is the variance of the Voronoi cell area distribution.

To quantify dynamical heterogeneity, we computed the four-point susceptibility $\chi_4$ following the framework of Lačević et al.(66). The overlap function $Q(t)$ is defined as:

$$Q(t) = \sum_{i=1}^{N}\sum_{j=1}^{N} w(|\mathbf{r}_i(0) - \mathbf{r}_j(t)|) \quad [16]$$

where $w(r)$ is a step function that selects particle pairs within a distance $a$, with $w(r) = 1$ for $r \leqslant a$ and 0 otherwise, and $a$ is the overlap radius. The four-point susceptibility is then given by:

$$\chi_4(t) = \frac{\beta V}{N^2}\left(\langle Q(t)^2 \rangle - \langle Q(t) \rangle^2\right) \quad [17]$$

where $\beta = 1/(k_B T)$, $V$ is the system volume, and $N$ is the total number of particles. The peak height $\chi_4^{max}$ and peak time $t_4^{max}$ were extracted from each $\chi_4$ curve as measures of the dynamical heterogeneity strength and its characteristic time scale.

## Acknowledgments

Bo Li acknowledges the support from taxpayers of China through the National Natural Science Foundation of China (Distinguished Young Scholars Funding, Overseas; Young Scientists Fund - Category C, Grant No. 32500692) and Wuhan University (Talents Startup Funding). Li Tian acknowledges the funding supported by the Wenzhou Municipal Bureau of Science and Technology under the Basic Public Welfare Research Project (Grant No. L2023003) and the Natural Science Foundation of Zhejiang Province under the Exploratory Youth Project (Grant No. LQ24A040006). We thank DeepSeek V3 for text polishing under the supervision of the authors.

## References

1. K. Binder, W. Kob, *Glassy materials and disordered solids: an introduction to their statistical mechanics* (World Scientific, 2005).

2. S. Sakka, J. D. Mackenzie, High pressure effects on glass. *Journal of Non-Crystalline Solids* **1**, 107–142 (1969).

3. N. V. Chandra Shekar, K. G. Rajan, Kinetics of pressure induced structural phase transitions—A review. *Bull Mater Sci* **24**, 1–21 (2001).

4. V. V. Brazhkin, A. G. Lyapin, High-pressure phase transformations in liquids and amorphous solids. *J. Phys.: Condens. Matter* **15**, 6059–6084 (2003).

5. C. Sonneville, *et al.*, Polyamorphic transitions in silica glass. *Journal of Non-Crystalline Solids* **382**, 133–136 (2013).

6. B. Haberl, *et al.*, Pressure-driven density match nucleates metastable r8 phases from amorphous Si and Ge. *Materials Today* **89**, 140–149 (2025).

7. P. W. Bridgman, I. Šimon, Effects of Very High Pressures on Glass. *Journal of Applied Physics* **24**, 405–413 (1953).

8. S. K. Deb, M. Wilding, M. Somayazulu, P. F. McMillan, Pressure-induced amorphization and an amorphous–amorphous transition in densified porous silicon. *Nature* **414**, 528–530 (2001).

9. J. Z. Jiang, *et al.*, Crystallization in Zr41.2Ti13.8Cu12.5Ni10Be22.5 bulk metallic glass under pressure. *Applied Physics Letters* **77**, 3553–3555 (2000).

10. S. Tsuneyuki, Y. Matsui, H. Aoki, M. Tsukada, New pressure-induced structural transformations in silica obtained by computer simulation. *Nature* **339**, 209–211 (1989).

11. J. S. Tse, D. D. Klug, J. A. Ripmeester, S. Desgreniers, K. Lagarec, The role of non-deformable units in pressure-induced reversible amorphization of clathrasils. *Nature* **369**, 724–727 (1994).

12. E. M. Stolper, T. J. Ahrens, On the nature of pressure‐induced coordination changes in silicate melts and glasses. *Geophysical Research Letters* **14**, 1231–1233 (1987).

13. A. Zeidler, *et al.*, High-Pressure Transformation of SiO 2 Glass from a Tetrahedral to an Octahedral Network: A Joint Approach Using Neutron Diffraction and Molecular Dynamics. *Phys. Rev. Lett.* **113**, 135501 (2014).

14. L. Stixrude, M. S. T. Bukowinski, Atomic structure of SiO 2 glass and its response to pressure. *Phys. Rev. B* **44**, 2523–2534 (1991).

15. K. Trachenko, M. T. Dove, Densification of silica glass under pressure. *J. Phys.: Condens. Matter* **14**, 7449–7459 (2002).

16. Y. Kono, *et al.*, Ultrahigh-pressure polyamorphism in $GeO_2$ glass with coordination number >6. *Proc. Natl. Acad. Sci. U.S.A.* **113**, 3436–3441 (2016).

17. C. Prescher, *et al.*, Beyond sixfold coordinated Si in $SiO_2$ glass at ultrahigh pressures. *Proc. Natl. Acad. Sci. U.S.A.* **114**, 10041–10046 (2017).

18. Q. Williams, R. Jeanloz, Spectroscopic Evidence for Pressure-Induced Coordination Changes in Silicate Glasses and Melts. *Science* **239**, 902–905 (1988).

19. P. Richet, Superheating, melting and vitrification through decompression of high-pressure minerals. *Nature* **331**, 56–58 (1988).

20. H. J. Jin, X. J. Gu, P. Wen, L. B. Wang, K. Lu, Pressure effect on the structural relaxation and glass transition in metallic glasses. *Acta Materialia* **51**, 6219–6231 (2003).

21. R. Meister, E. C. Robertson, R. W. Werre, R. Raspet, Elastic moduli of rock glasses under pressure to 8 kilobars and geophysical implications. *J. Geophys. Res.* **85**, 6461–6470 (1980).

22. Arndt, D. Stöffler, Anomalous changes in some properties of silica glass densified at very high pressures. **10**, 117–124 (1969).

23. H. M. Cohen, R. Roy, Effects of Ultra high Pressures on Glass. *Journal of the American Ceramic Society* **44**, 523–524 (1961).

24. M. Vaccari, *et al.*, Structural changes in amorphous GeS 2 at high pressure. *Phys. Rev. B* **81**, 014205 (2010).

25. P. J. Lu (陸述義), D. A. Weitz, Colloidal Particles: Crystals, Glasses, and Gels. *Annu. Rev. Condens. Matter Phys.* **4**, 217–233 (2013).

26. D. Frenkel, Playing Tricks with Designer "Atoms." *Science* **296**, 65–66 (2002).

27. D. Frenkel, Colloidal Encounters: A Matter of Attraction. *Science* **314**, 768–769 (2006).

28. J. C. Crocker, D. G. Grier, Methods of Digital Video Microscopy for Colloidal Studies. *Journal of Colloid and Interface Science* **179**, 298–310 (1996).

29. K. Hima Nagamanasa, S. Gokhale, A. K. Sood, R. Ganapathy, Direct measurements of growing amorphous order and non-monotonic dynamic correlations in a colloidal glass-former. *Nature Phys* **11**, 403–408 (2015).

30. B. Li, K. Lou, W. Kob, S. Granick, Anatomy of cage formation in a two-dimensional glass-forming liquid. *Nature* **587**, 225–229 (2020).

31. E. R. Weeks, J. C. Crocker, A. C. Levitt, A. Schofield, D. A. Weitz, Three-Dimensional Direct Imaging of Structural Relaxation Near the Colloidal Glass Transition. *Science* **287**, 627–631 (2000).

32. N. Singh, A. K. Sood, R. Ganapathy, Cooperatively rearranging regions change shape near the mode-coupling crossover for colloidal liquids on a sphere. *Nat Commun* **11**, 4967 (2020).

33. B. Zhang, X. Cheng, Structures and Dynamics of Glass-Forming Colloidal Liquids under Spherical Confinement. *Phys. Rev. Lett.* **116**, 098302 (2016).

34. G. L. Hunter, K. V. Edmond, E. R. Weeks, Boundary Mobility Controls Glassiness in Confined Colloidal Liquids. *Phys. Rev. Lett.* **112**, 218302 (2014).

35. S. Lang, *et al.*, Glass Transition in Confined Geometry. *Phys. Rev. Lett.* **105**, 125701 (2010).

36. B. De Nijs, *et al.*, Entropy-driven formation of large icosahedral colloidal clusters by spherical confinement. *Nature Mater* **14**, 56–60 (2015).

37. W. T. M. Irvine, V. Vitelli, P. M. Chaikin, Pleats in crystals on curved surfaces. *Nature* **468**, 947–951 (2010).

38. G. Meng, J. Paulose, D. R. Nelson, V. N. Manoharan, Elastic Instability of a Crystal Growing on a Curved Surface. *Science* **343**, 634–637 (2014).

39. R. E. Guerra, C. P. Kelleher, A. D. Hollingsworth, P. M. Chaikin, Freezing on a sphere. *Nature* **554**, 346–350 (2018).

40. C. F. Mbah, *et al.*, Early-stage bifurcation of crystallization in a sphere. *Nat Commun* **14**, 5299 (2023).

41. N. Singh, A. K. Sood, R. Ganapathy, Observation of two-step melting on a sphere. *Proc. Natl. Acad. Sci. U.S.A.* **119**, e2206470119 (2022).

42. B. V. R. Tata, P. S. Mohanty, M. C. Valsakumar, Glass Transition and Dynamical Heterogeneities in Charged Colloidal Suspensions under Pressure. *Phys. Rev. Lett.* **88**, 018302 (2001).

43. T. Voigtmann, W. C. K. Poon, Glasses under high pressure: a link to colloidal science? *J. Phys.: Condens. Matter* **18**, L465–L469 (2006).

44. A. Hasmy, S. Ispas, B. Hehlen, Percolation transitions in compressed SiO2 glasses. *Nature* **599**, 62–66 (2021).

45. Z. Zhang, Z. Xie, W. Kob, Symmetry transitions beyond the nanoscale in pressurized silica glass. *Proc. Natl. Acad. Sci. U.S.A.* **122**, e2524058122 (2025).

46. W. Kob, H. C. Andersen, Testing mode-coupling theory for a supercooled binary Lennard-Jones mixture. II. Intermediate scattering function and dynamic susceptibility. *Phys. Rev. E* **52**, 4134–4153 (1995).

47. A. Jayaraman, Diamond anvil cell and high-pressure physical investigations. *Rev. Mod. Phys.* **55**, 65–108 (1983).

48. S. Vivek, C. P. Kelleher, P. M. Chaikin, E. R. Weeks, Long-wavelength fluctuations and the glass transition in two dimensions and three dimensions. *Proc. Natl. Acad. Sci. U.S.A.* **114**, 1850–1855 (2017).

49. R. Abeyaratne, Lecture notes on the mechanics of elastic solids. MIT. (1987). [Accessed 3 January 2023].

50. R. Abeyaratne, Lecture Notes on The Mechanics of Elastic Solids Volume II: Continuum Mechanics. 428 (2012).

51. O. L. Anderson, J. E. Nafe, The bulk modulus-volume relationship for oxide compounds and related geophysical problems. *J. Geophys. Res.* **70**, 3951–3963 (1965).

52. T. Rouxel, H. Ji, T. Hammouda, A. Moréac, Poisson's Ratio and the Densification of Glass under High Pressure. *Phys. Rev. Lett.* **100**, 225501 (2008).

53. R. Ota, T. Yamate, N. Soga, M. Kunugi, Elastic properties of Ge Se glass under pressure. *Journal of Non-Crystalline Solids* **29**, 67–76 (1978).

54. N. Y. C. Lin, M. Bierbaum, P. Schall, J. P. Sethna, I. Cohen, Measuring nonlinear stresses generated by defects in 3D colloidal crystals. *Nature Mater* **15**, 1172–1176 (2016).

55. L. Berthier, *et al.*, Direct Experimental Evidence of a Growing Length Scale Accompanying the Glass Transition. *Science* **310**, 1797–1800 (2005).

56. L. Berthier, E. Flenner, G. Szamel, Comment on "Fickian Non-Gaussian Diffusion in Glass-Forming Liquids." *Phys. Rev. Lett.* **131**, 119801 (2023).

57. J. D. Stevenson, J. Schmalian, P. G. Wolynes, The shapes of cooperatively rearranging regions in glass-forming liquids. *Nature Phys* **2**, 268–274 (2006).

58. Z. Zhang, P. J. Yunker, P. Habdas, A. G. Yodh, Cooperative Rearrangement Regions and Dynamical Heterogeneities in Colloidal Glasses with Attractive Versus Repulsive Interactions. *Phys. Rev. Lett.* **107**, 208303 (2011).

59. B. Abou, D. Bonn, J. Meunier, Aging dynamics in a colloidal glass. *Phys. Rev. E* **64**, 021510 (2001).

60. P. G. Debenedetti, F. H. Stillinger, Supercooled liquids and the glass transition. *Nature* **410**, 259–267 (2001).

61. Z. Zhang, W. Kob, Revealing the three-dimensional structure of liquids using four-point correlation functions. *Proc. Natl. Acad. Sci. U.S.A.* **117**, 14032–14037 (2020).

62. W. Kob, S. Roldán-Vargas, L. Berthier, Non-monotonic temperature evolution of dynamic correlations in glass-forming liquids. *Nature Phys* **8**, 164–167 (2012).

63. D. Kaya, N. L. Green, C. E. Maloney, M. F. Islam, Normal Modes and Density of States of Disordered Colloidal Solids. *Science* **329**, 656–658 (2010).

64. B. Li, D. Zhou, Y. Han, Assembly and phase transitions of colloidal crystals. *Nat Rev Mater* **1**, 15011 (2016).

65. L. Tian, “Phase behaviours of colloidal systems with critical Casimir forces.” (2022).

66. N. Lačević, F. W. Starr, T. B. Schrøder, S. C. Glotzer, Spatially heterogeneous dynamics investigated via a time-dependent four-point density correlation function. *The Journal of Chemical Physics* **119**, 7372–7387 (2003).

## Figures and legends

**Figure 1. Structural and dynamic characterization of a monolayer colloidal glass on the inner wall of a capillary. (A)** Schematic of the experimental setup. Bidisperse silica particles sediment onto the inner wall of a cylindrical capillary under gravity, forming a monolayer colloidal glass. The observation zone at the bottom center of the capillary is marked by the green dashed region, while the particles on both sidewalls exert a lateral confining pressure on the central region (red dashed region), arising from the tangential component of gravity along the curved wall. **(B)** Magnified view of the region of interest (ROI) and the lateral confinement. The ROI is outlined by the green rectangle; red arrows indicate the direction of the confining pressure from the sidewalls. **(C)** Three-dimensional reconstructions of particles on the capillary wall (side view). Color represents the particle height (*z* coordinate), with $z = 0$ set at the bottom center of the tube. Top: curvature $\kappa = 6.67\ \mathrm{mm}^{-1}$ (0.3 mm capillary); bottom: $\kappa = 3.33\ \mathrm{mm}^{-1}$ (0.6 mm capillary). **(D–F)** *MSD* on double-logarithmic scales. **(D)** $\kappa = 6.67\ \mathrm{mm}^{-1}$, **(E)** $\kappa = 3.33\ \mathrm{mm}^{-1}$, **(F)** two-dimensional reference sample. Different colors and symbols denote different $P$ in (**D, E**) and different $\varphi$ in (**F**) (see legends). *MSD* plateaus appear for all pressures in (D, E), but only for $\varphi = 0.83$ in (F). **(G–I)** Self-intermediate scattering functions $F_s(q,t)$. **(G)** $\kappa = 6.67\ \mathrm{mm}^{-1}$, **(H)** $\kappa = 3.33\ \mathrm{mm}^{-1}$, **(I)** two-dimensional reference sample. The decay of $F_s(q,t)$ slows down with increasing pressure (**G, H**) and volume fraction (**I**), consistent with the *MSD* results.

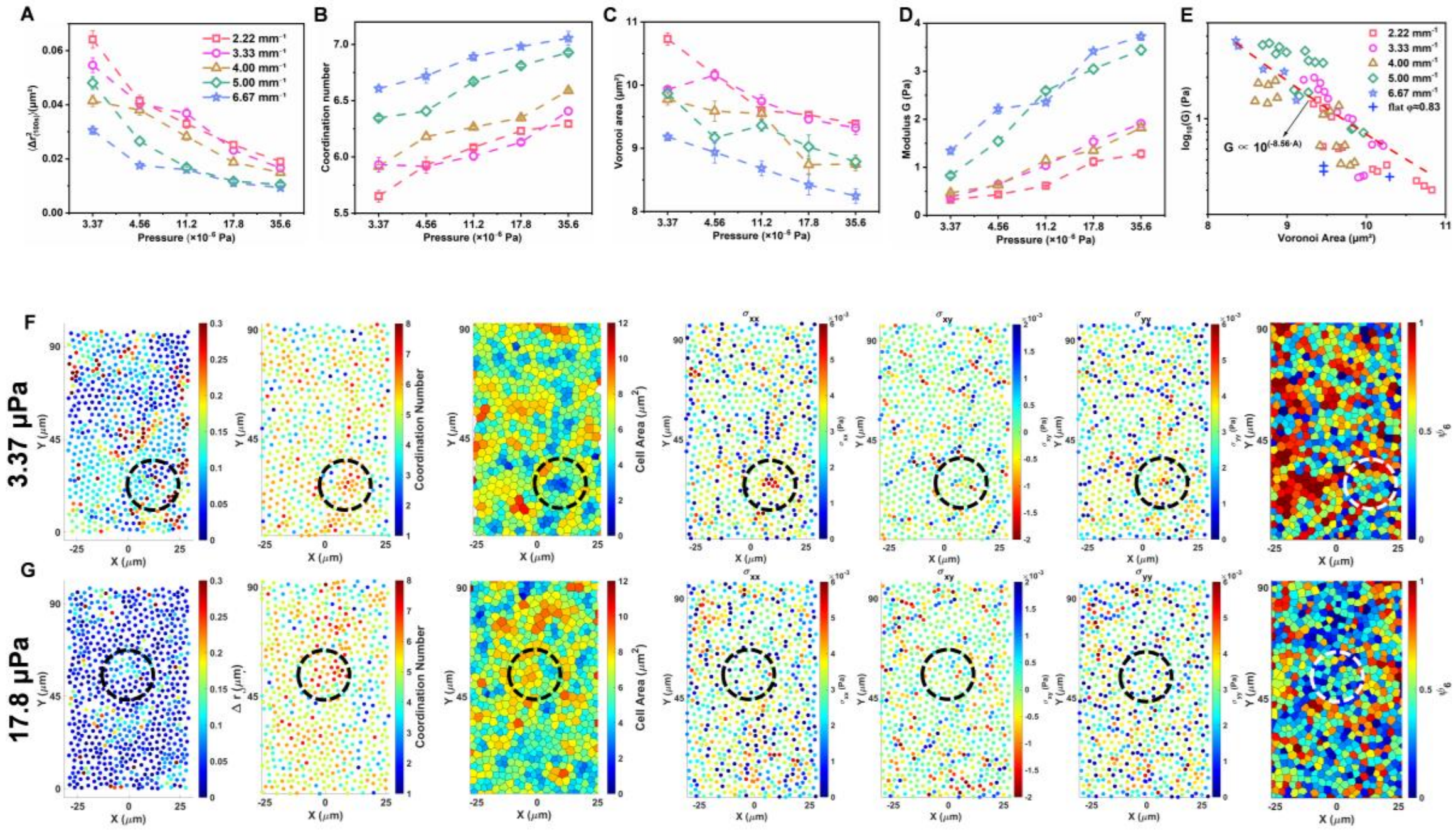

**Figure 2. Pressure-dependent structure, dynamics, and structural–mechanical coupling.** **(A)** *MSD* at $t$ = 100 s as a function of pressure for five curvatures. **(B)** Coordination number as a function of pressure for five curvatures. **(C)** Voronoi cell area as a function of pressure. **(D)** Shear modulus $G$ as a function of pressure. Legends and error bars (SD) are the same as in **(A)**. **(E)** $log_{10}(G)$ as a function of Voronoi cell area. Different symbols denote different curvatures; the flat reference sample ($\varphi$ = 0.83) is included for comparison. The red dashed line represents an exponential fit: $log_{10}(G) = 2.78 - 8.56 \cdot A$. **(F)** $\kappa$ = 6.67 mm$^{-1}$, $P$ = 3.37 μPa. From left to right: particle displacement ($\Delta t$ ≈13.9 s), coordination number, Voronoi cell area, stress tensor components $\sigma_{xx}$, $\sigma_{yy}$, $\sigma_{xy}$ (calculated using the SALSA stress formula), and $\psi_6$-colored Voronoi cells. The horizontal and vertical axes are the projected coordinates of particles onto the two-dimensional plane (unit: μm). **(G)** $\kappa$ = 6.67 mm$^{-1}$, $P$ = 17.8 μPa. The arrangement and axes are the same as in **(F)**.

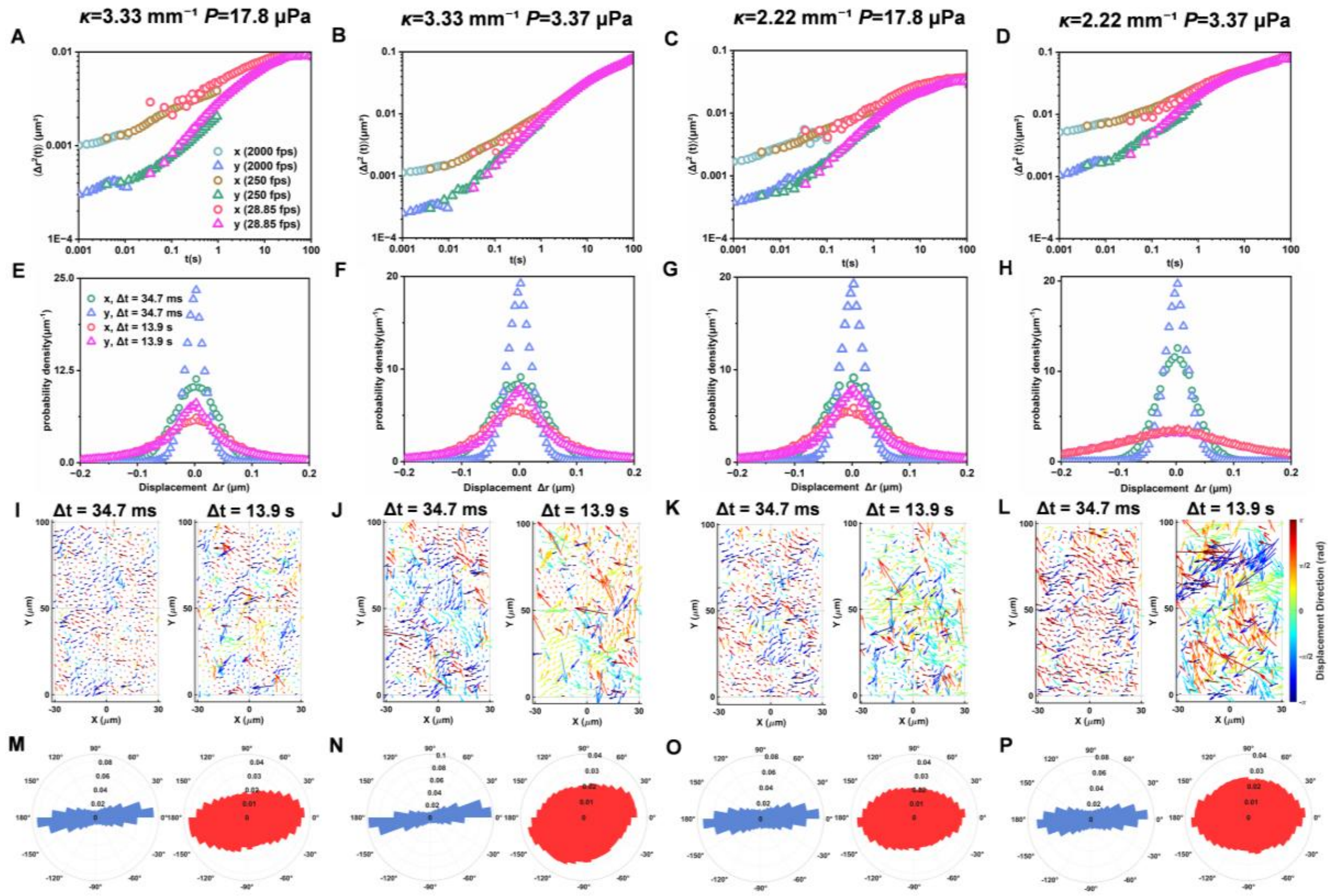

**Figure 3. Comparison of dynamics in the X and Y directions under different curvatures and pressures. (A–D)** *MSD* curves. Different colors denote different frame rates (see legend); open circles and open triangles represent the x- and y-directions, respectively. The four conditions are: **(A)** $\kappa$ = 3.33 mm$^{-1}$, $P$ = 1.78 × 10$^{-5}$ Pa; **(B)** $\kappa$ = 3.33 mm$^{-1}$, $P$ = 3.37 × 10$^{-6}$ Pa; **(C)** $\kappa$ = 2.22 mm$^{-1}$, $P$ = 1.78 × 10$^{-5}$ Pa; **(D)** $\kappa$ = 2.22 mm$^{-1}$, $P$ = 3.37 × 10$^{-6}$ Pa. **(E–H)** Displacement probability density distributions (corresponding to conditions **A–D**). Symbols are the same as in **(A–D)**. **(I–L)** Displacement arrow maps (corresponding to conditions **A–D**). Each panel consists of two subfigures: left: $\Delta t$ ≈ 35 ms; right: $\Delta$t ≈ 13.9 s. Arrow length represents displacement magnitude; color indicates displacement direction angle (−π to π, see color bar). The horizontal and vertical axes are the projected coordinates of particles onto the two-dimensional plane (unit: μm). **(M–P)** Rose diagrams of displacement direction (corresponding to conditions **A–D**). Left: $\Delta$t ≈ 34.7 ms; right: $\Delta$t ≈ 13.9 s. Radial length indicates the probability density in each angular bin.

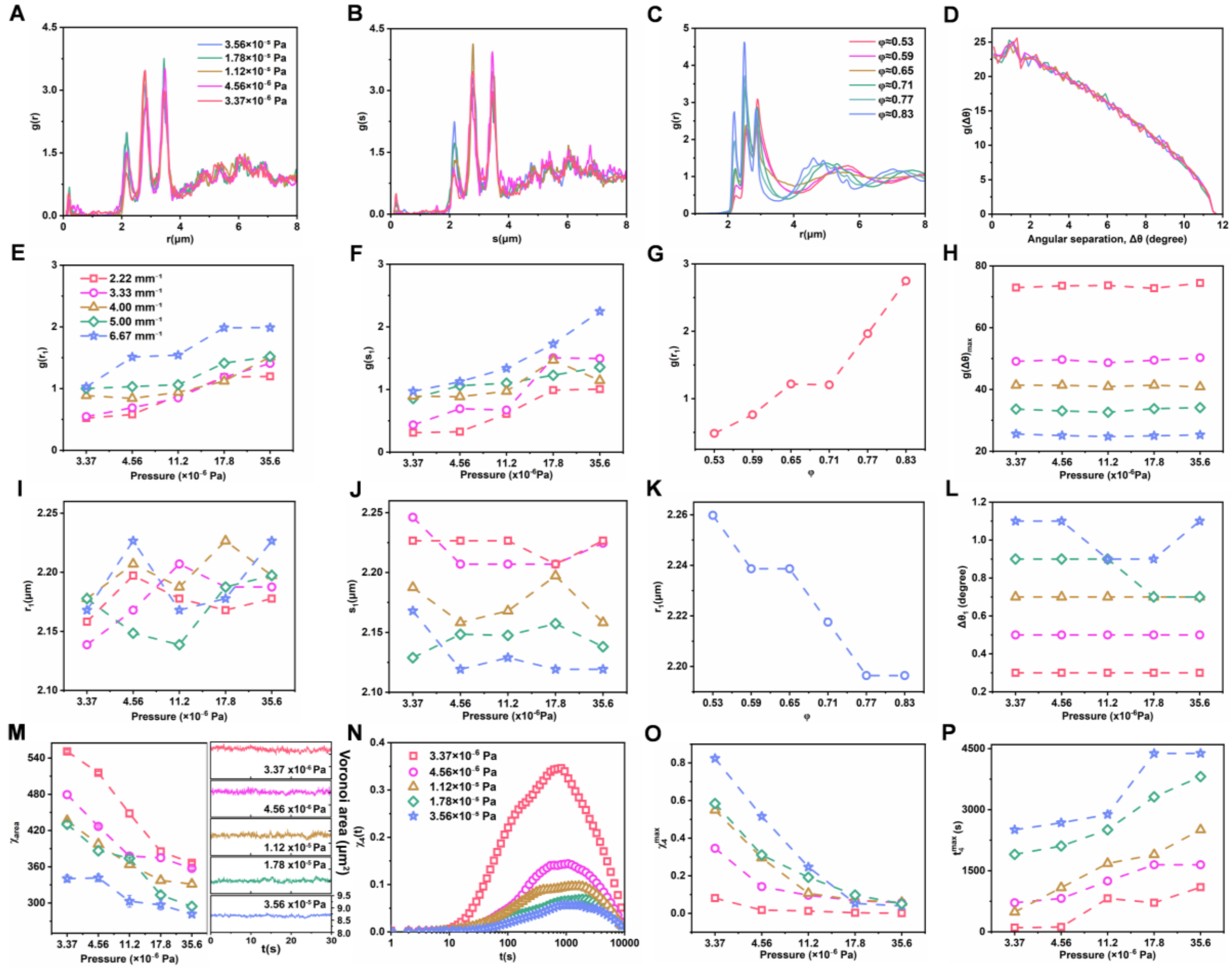

**Figure 4. Structural correlations, area variation, and dynamical heterogeneity. (A)** $g(r)$ as a function of distance $r$ at $\kappa$ = 6.67 mm$^{-1}$ for five pressures (from 3.37 × 10$^{-6}$ to 3.56 × 10$^{-5}$ Pa). **(B)** $g(s)$ as a function of geodesic distance $s$ at $\kappa$ = 6.67 mm$^{-1}$ for five pressures. **(C)** $g(r)$ for the two-dimensional reference sample at six volume fractions (from $\varphi$ = 0.53 to 0.83, see legend). **(D)** Angular pair distribution function $g(\Delta\theta)$ at $\kappa$ = 6.67 mm$^{-1}$ for five pressures. $\Delta\theta$ is the angular separation between two particles along the circumferential direction of the tube wall. **(E–L)** First peak height and first peak position as functions of pressure or volume fraction: **(E)** peak height of $g(r)$ vs $P$; **(F)** peak height of $g(s)$ vs $P$; **(G)** peak height of $g(r)$ vs $\varphi$ (flat cell); **(H)** peak height of $g(\Delta\theta)$ vs $P$; **(I)** peak position of $g(r)$ vs $P$; **(J)** peak position of $g(s)$ vs $P$; **(K)** peak position of $g(r)$ vs $\varphi$ ; **(L)** peak position of $g(\Delta\theta)$ vs $P$. **(M) Left:** Area compressibility $\chi_{\text{area}}$ as a function of pressure, where $\chi_{\text{area}} = A \cdot \text{Var}(a) / \langle a\rangle^2$, quantifying the relative variation of Voronoi cell areas with time. Error bars represent *SD*. **Right:** Voronoi cell area as a function of time at $\kappa$ = 6.67 mm$^{-1}$ for five pressures (from low to high, top to bottom). Each curve represents the mean Voronoi cell area averaged over all particles within the corresponding frame. **(N)** Time dependence of the four-point susceptibility $\chi_4(t)$ at curvature $\kappa$ = 6.67 mm$^{-1}$ for different pressures. As pressure decreases, the peak of $\chi_4(t)$ monotonically grows and shifts to shorter times. **(O)** Peak height $\chi_4^{max}$ as a function of pressure for five curvatures. **(P)** Peak time $t_4^{max}$ as a function of pressure for five curvatures. For panels (**E, F, H, I, J, L, M, O, P**), different colors and symbols denote different curvatures, sharing the same legend (see **E**).